\documentclass[english,preprint]{revtex4}
\usepackage[T1]{fontenc}
\usepackage[latin1]{inputenc}
\usepackage{amsmath}
\usepackage{amssymb}

\makeatletter
\usepackage{graphicx}

\usepackage{setspace}

\usepackage{babel}
\makeatother
\begin{document}
\def\figurename{FIG.}
\def\abstractname{}
\title{Non summable Borel $\Phi^{4}$ theory in zero dimensions and the
Generalized Borel Transform}

\author{M. Marucho}

\affiliation{Department of Chemistry, University of
Houston, Houston, TX 77204-5641, USA.}
\author{\emph {marucho@rn.chem.uh.edu}}
\begin{abstract}
A new technique named Generalized Borel Transform (GBT) is applied
to the generating functional of the $\Phi^{4}$ theory in zero dimensions
with degenerate minima. The analytical solution of this function,
obtained in the non perturbative regime, is compared with those estimations
predicted by large order perturbation theory. It was established that
the GBT is a very efficient technique to capture these contributions.
On the other hand, renormalons associated to the resummation of those
perturbative series were not found to be the genuine source of the
non perturbative contributions of this model.
\end{abstract}
\maketitle

Substantial improvement of experimental data has recently demanded theorists
to give more accurate estimates since the finite order perturbative
predictions have uncertainties comparable with experimental errors. This has
renewed the interest in computing non perturbative contributions from the
resummation of series. In Quantum Field theory non summable Borel, the
problem of obtaining these contributions has been largely studied in the
literature.$^1$ Among the different methods proposed to tackle this problem,
the Conventional Borel Transform (CBT) of the weak coupling expansion$^2$
and its modified version (MBT)$^{3,4}$ have become popular tools for the
reconstruction of the physical quantities predicted by these theories.
Basically, these techniques require the exact knowledge of the large order
behavior of the perturbative series in the coupling parameter. In spite of
the lack of these series, which are divergent,$^5$ one can make use of their
coefficients to define other expansion in powers of a new variable named
Borel and in such a way that it is convergent. Then, the resummation of the
initial asymptotic series is formally recovered by integrating the
convergent series on the Borel variable. In this approach, the non
perturbative contributions are estimated from the ambiguity appearing in the
integration, where singularities (poles and/or branches) called IR
renormalons$^6$ localized on the positive real semiaxis of the Borel complex
plane invalidate these proposals. On the other hand, an alternative
development, called the Generalized Borel Transform (GBT),$^{7-9}$ was
recently presented which has proved to be potentially useful in overcoming
those difficulties. It avoids the use of series by introducing auxiliary
parameters having no physical meaning which are introduced for the sole
purpose of making the calculation analytically tractable. These parameters
together with the analytic properties of the GBT are the fundamental tools
needed to obtain analytical solutions of parametric integrals like
Laplace-Fourier-Mellin transforms (LFMT) for all the range of its parameter.
Therefore, it is extremely useful to study non-perturbative regimes. In
addition, it involves simple mathematics which requires basic calculus such
as the evaluation of derivative, indefinite integrals and limits. In fact,
this technique has already been successfully applied to different areas of
Physics such as Quantum Field Theory,$^7$ Polymer Physics$^8$ and Quantum
Mechanics.$^9$

In connection with the techniques previously mentioned, it is certainly
instructive to study exactly solvable models so that the efficiency and
precision of the different proposals can be quantitatively checked. This
analysis may help in defining criteria for selecting the appropriate
approach to be used in more realistic models. In this sense, it is largely
accepted that the analysis of zero-dimensional models can provide some
insight on the behavior of Green's functions in Field theories and
Statistical systems.$^{10}$ In particular, the understanding of some aspects
of zero-dimensional Field Theories could leave trails about the non
perturbative behavior of these theories in higher dimensions. For instance,
classical actions with degenerate isolated minima play a fundamental role in
studying processes governing the decay of metastable atomic and nuclear
states as well as the transition of overheated or undercooled thermodynamic
phases to a stable equilibrium phase.

One of the most popular model of this kind is the well-known
zero-dimensional $\Phi ^4$ theory with a degenerate minima$^1$ due to its
purely non perturbative feature. The analysis of this model is often based
on the study of its generating function, starting point for computing
moments and Green's functions. Moreover, at present the exact solution is
known from the numerical computation of its integral representation

\begin{equation}
Z\left(g\right)=\int_{-\infty}^{+\infty}dx\exp\left[-x^{2}\left(1-gx%
\right)^{2}/2\right],\qquad g\geq0.  \label{fungen}
\end{equation}

Therefore, the main purpose of this paper consists in discussing the use and
to remark the advantages of the different approximate computing tools
aforementioned to obtain an analytical approximate solution of the
generating function (\ref{fungen}) in the non perturbative regime.

Thus, the analysis of this model is initially done on the context of
Perturbative Field Theory in which non perturbative contributions are
estimated from the solution of the generating functional (\ref{fungen}) in
powers of $g.$ It is easily obtained by rewriting the integrand in Eq.(\ref
{fungen}) as a product of two exponential functions as follows

\[
Z\left(g\right)=\int_{-\infty}^{+\infty}dx\exp\left[-x^{2}/2\right]\exp%
\left[gx^{3}-g^{2}x^{4}/2\right]. 
\]

Then, the expansion of the second exponential in powers of its argument
leads to the following formal solution of this model

\begin{equation}
Z_{P}\left(g\right)=\sqrt{2\pi}+\sqrt{2}\sum_{n=0}^{\infty}\frac{%
\Gamma\left(2n+5/2\right)}{\Gamma\left(n+2\right)}\left(8g^{2}\right)^{n+1},%
\quad g\rightarrow0^{+}.  \label{exactseries}
\end{equation}

Even though this solution is exact, it is plagued of difficulties and
consequently it does not represent a suitable analytical solution. Indeed,
the series (\ref{exactseries}) is asymptotic and factorial divergent.
Basically, it means that the contributions of the terms of the series
decrease with increasing $n$ for $n<n_{min}$ and increase for $n>n_{min}$.
The value of $n_{min}=N_o\left( g\right) $ corresponds to the term with the
minimal contribution to the series. Thus, the best approximation provided by
expression (\ref{exactseries}) is obtained by truncating the series at $%
n_{min}$. Given that it depends on the value of the parameter $g$, then the
accuracy of the result will also depend on it. For example, small values of $%
g$ provide a large $N_o\left( g\right) $ and consequently there are
sufficient amounts of terms contained in this series so that a good
approximation of the exact solution is obtained. In contrast to this, values
of the parameter $g$ going away from the perturbative regime reduce the
value of $N_o\left( g\right) $ and the series involves just a few terms.
Thus, it is significantly off from the exact one. In addition to this
trouble, their coefficients do not alternate in sign avoiding the
cancelation of the consecutive terms in the series, which makes it worse the
divergent behavior.

In some cases, the aforementioned difficulties can be minimized, or
hopefully eliminated, by applying standard approaches specifically designed
to deal with this class of series. In this model, it is clear that the use
of the exact expression of their coefficients would simply recover the
starting expression for the generating function given by the integral
representation (\ref{fungen}) whose analytical solution is unknown. However,
an approximate analytical solution for this function can be obtained from
Eq. (\ref{exactseries}) by using the following asymptotic behavior for their
coefficients

\begin{equation}
\frac{\Gamma \left( 2n+5/2\right) }{\Gamma \left( n+2\right) }\sim 4^n\Gamma
\left( n+1\right) 2\sqrt{2/\pi }  \label{firstapprox}
\end{equation}

in which case the resulting approximate generating function becomes a
factorial divergent series

\begin{equation}
Z_P^{as}\left( g\right) =\sqrt{2\pi }+\frac 1{\sqrt{\pi }}\sum_{n=0}^\infty
\Gamma \left( n+1\right) \left( 32g^2\right) ^{n+1},\quad g\rightarrow 0^{+}.
\label{firstasymp}
\end{equation}

which can be resumed by applying the Conventional Borel Transform as
follows. The Borel sum of the generating function is obtained from
expression (\ref{firstasymp}) by using the integral representation of the
Gamma function.$^{11}$ Afterwards, the sum and the integral operations
presented there are formally interchanged to get

\begin{equation}
Z_{Borel}\left(g\right)=\sqrt{2\pi}+\int_{0}^{+\infty}dt\exp\left[-t\right]B%
\left(t\right),  \label{borelsum}
\end{equation}

where $B\left(t\right)$ is the Borel Transform defined by

\begin{equation}
B\left(t\right)\equiv\frac{32g^{2}}{\sqrt{\pi}}\sum_{n=0}^{\infty}%
\left(32tg^{2}\right)^{n}=\frac{32g^{2}}{\sqrt{\pi}}\frac{1}{1-32tg^{2}}
\label{boreltransf}
\end{equation}

Since the last expression is valid only for $32tg^2<1$, the integral in Eq.(%
\ref{borelsum}) can be computed whenever an analytical continuation of its
integrand to the Borel complex plane is available. In doing so, it appears a
singularity at $t=1/32g^2$. As a consequence, the resummation becomes
invalid (non summable Borel). Anyway, a meaning to the ill-defined integral (%
\ref{borelsum}) is often given by adopting a regularization prescription.
Nevertheless, this generates an ambiguity in its evaluation since the result
will depend on the particular prescription taken. If the Cauchy Principal
value is used, the pole is avoided by deforming the contour of integration
along two semicircles centered at the pole above and below the real axis
respectively. Hence,$^{6,12}$

\begin{equation}
Z_{Borel}\left(g\right)=\sqrt{2\pi}+{\frac{32g^{2}}{\sqrt{\pi}}}%
PV\int_{0}^{+\infty}dt{\frac{\exp\left[-t\right]}{1-32tg^{2}}}\pm i\sqrt{\pi}%
\exp\left[-1/32g^{2}\right]  \label{firstregborel}
\end{equation}

where the last term comes from the residue evaluated at the pole.

From this definition, the estimation of the non perturbative contributions
is immediate. The generating function defined initially by equation (\ref
{fungen}) is analytic in the whole complex plane of the coupling parameter $%
g $. Then, the non perturbative contributions which had been excluded in the
computation, must cancel those non analytical contributions present in the
expression (\ref{firstregborel}) in order to preserve this property. In this
sense, the non analytical structure of the Borel transform is assumed to
contain information about the origin of the main non perturbative
contributions of the model, leaving aside the ones unrelated to renormalons.

Therefore, within this technique one ends with an approximate solution for
the generating function in the form$^3$

\begin{equation}
Z_{Borel}\left( g\right) =\sqrt{2\pi }+\frac{32g^2}{\sqrt{\pi }}\exp \left[
-\frac 1{32g^2}\right] Ei\left( \frac 1{32g^2}\right) +C\exp \left[
-1/32g^2\right]   \label{regborel22}
\end{equation}
where $Ei\left( z\right) $ is the exponential integral function$^{11}$ and $C
$ is a constant to be adjusted from {}``experiment''. In this model it means
from the numerical exact solution (\ref{fungen}).

The numerical evaluation of this result is presented in Fig. 1. This plot
clearly shows that the prediction coming from the large order perturbative
theory is not good for the non perturbative regime. Therefore, the non
analyticity of the Borel sum (IR renormalons) is not the true source of the
main non perturbative contributions of this model.

In more realistic theories the exact expression of the coefficients of the
series is otherwise rarely known.$^1$ However, one can analyze other
asymptotic expansion of the coefficients in this model which converge more
rapidly to the exact one. Concretely, one of them reads

\begin{equation}
Z_{P}^{as}\left(g\right)=\sqrt{2\pi}+\frac{1}{\sqrt{\pi}}\sum_{n=0}^{\infty}%
\frac{\Gamma\left(n+3/2\right)\left(32g^{2}\right)^{n+1}}{%
\left(n+1\right)^{1/2}},\quad g\rightarrow0^{+}.  \label{asympseries}
\end{equation}

By applying the previously explained Borel approach to this series, the
Borel Transform adopts the form

\begin{equation}
B\left(t\right)\equiv\frac{t^{-1/2}}{\sqrt{\pi}}\sum_{n=0}^{\infty}\frac{%
\left(32tg^{2}\right)^{n}}{\left(n+1\right)^{1/2}}  \label{boreltransf2}
\end{equation}

This series converges to the Polylog function$^{11}$ $L_{1/2}\left(
32tg^2\right) $ in the interval $\left[ 0,1/32g^2\right] .$ Explicitly, the
expression (\ref{borelsum}) becomes

\begin{equation}
Z_{Borel}\left( g\right) =\sqrt{2\pi }+\frac 1{\sqrt{\pi }}\int_0^{+\infty
}dt\exp \left[ -t\right] t^{-1/2}L_{1/2}\left( 32tg^2\right) 
\label{borelsum2}
\end{equation}
where right now the integrand in Eq.(\ref{borelsum2}) must be analytically
continuated to $\left| 32tg^2\right| \geq 1$. As a consequence, it acquires
a cut on the positive real semiaxis in the interval $\left[ 1/32g^2,\infty
\right] $ due to the well-known Polylog function analytical properties{.}$%
^{11}$ As expected, the resummation becomes invalid again. In this case, the
non analyticity is avoided by deforming the contour of integration along two
parallel lines slightly above and below the real axis. Hence, it reads$^{12}$

\begin{equation}
Z_{Borel}\left(g\right)=\sqrt{2\pi}+\frac{1}{\sqrt{\pi}}\int_{0}^{+\infty}dt%
\exp\left[-t\right]t^{-1/2}Re\left\{ L_{1/2}\left(32tg^{2}\right)\right\}
\label{regborel}
\end{equation}

\[
\pm\frac{i}{\sqrt{\pi}}\int_{1/32g^{2+}}^{+\infty}dt\exp\left[-t%
\right]t^{-1/2}Im\left\{ L_{1/2}\left(32tg^{2}\right)\right\} 
\]
with

\[
Im\left\{ L_{1/2}\left( 32tg^2\right) \right\} =-\frac{\sqrt{\pi }}{\sqrt{%
\ln \left( 32tg^2\right) }}.
\]
being the non analytical part giving an estimation of the non perturbative
contributions. Therefore, the suitable approximate solution for the
generating function has the form$^3$

\begin{equation}
Z_{Borel}\left(g\right)=\frac{1}{\sqrt{\pi}}\int_{0}^{+\infty}dt\exp\left[-t%
\right]t^{-1/2}Re\left\{ L_{1/2}\left(32tg^{2}\right)\right\}
\label{regborel1}
\end{equation}

\[
+C\int_{1/32g^{2}}^{+\infty}dt\frac{\exp\left[-t\right]t^{-1/2}}{\sqrt{%
\ln\left(32tg^{2}\right)}}, 
\]

Unfortunately, the analytical solution of these integrals is unknown. The
numerical evaluation of this result (Fig. 1) shows that the approximation (%
\ref{regborel1}) improves the prediction of non perturbative contributions
when it is compared with the first approach (\ref{regborel22}). In spite of
this improvement, the expression (\ref{regborel1}) still does not provide a
good estimation of the generating function (\ref{fungen}) in the non
perturbative regime.

Setting aside the difficulties present in the Borel resummation technique,
the computation of the generating function by using the GBT is now
presented. In doing so, it is necessary to rewrite the expression (\ref
{fungen}) in terms of a Laplace transform. It is achieved as follows.
Firstly, the variable of integration $x$ is replaced by $u/g$ and a new
coupling parameter $a\equiv1/\left[2g^{2}\right]$ is defined. Secondly, the
integration on the whole real axis is performed on each semiaxis, positive
and negative, separately. In addition, the last integral is rewritten on the
positive semiaxis by changing variables $u=-v$. As a consequence, the
expression (\ref{fungen}) becomes

\begin{equation}
Z\left(a\right)=\sqrt{2a}\left\{
\int_{0}^{+\infty}du\exp\left[-au^{2}\left(1+u\right)^{2}\right]+\int_{0}^{+%
\infty}dv\exp\left[-av^{2}\left(1-v\right)^{2}\right]\right\} .
\label{interexp}
\end{equation}

Then, the second integral in Eq.(\ref{interexp}) is divided into $%
\int_{0}^{1}$ plus $\int_{1}^{+\infty}$ so that the change of variables $%
w=1-v$ transforms $\int_{1}^{+\infty}$ into the first integral in Eq.(\ref
{interexp}). Finally, this is written as a Laplace transform by changing
variables $y=u^{2}\left(1+u\right)^{2}$ to get

\begin{equation}
Z\left(a\right)=\sqrt{2a}\left\{ \int_{0}^{+\infty}dy\frac{%
\exp\left(-ay\right)}{\sqrt{y}\sqrt{1+4\sqrt{y}}}+\int_{0}^{1}dv\exp%
\left[-av^{2}\left(1-v\right)^{2}\right]\right\} .  \label{newexpgenfun}
\end{equation}

Observe that this new expression for the generating function presents
important advantages. A simple analysis of the behavior of both integrals in
Eq.(\ref{newexpgenfun}) allows one to identify what is the source of the non
perturbative contributions. In fact, they show that, in the limit $%
g\rightarrow\infty^{+}$ $\left(a\rightarrow0^{+}\right),$ the main non
perturbative contribution comes from the Laplace transform. Furthermore, it
can be analytically computed by applying the GBT technique which is
described below.

Basically this method consists of introducing two auxiliary functions, $%
S\left(a,n\right)$ and $B_{\lambda}\left(s,n\right)$(the Generalized Borel
Transform) which depend on auxiliary parameters called $n$ and $\lambda$.
These parameters have no physical meaning and are introduced for the sole
purpose of helping in the computation of an explicit mathematical expression
for the following Laplace transform

\begin{equation}
S\left(a\right)=\int_{0}^{+\infty}duH\left(u\right)\exp\left(-au\right),%
\qquad a>0.  \label{laplace}
\end{equation}

In doing so, $S\left(a,n\right)$ is defined in terms of $S\left(a\right)$ by
the formula

\begin{equation}
S\left(a,n\right)\equiv\left(-\right)^{n}\frac{\partial^{n}}{\partial a^{n}}%
S\left(a\right)=\int_{0}^{\infty}u^{n}H\left(u\right)\exp\left(-au\right)du,%
\qquad n\geq0.  \label{int}
\end{equation}
which can be inverted to get

\begin{equation}
S\left(a\right)=\left(-\right)^{n}\underbrace{\int da\cdots\int da}%
_{n}S\left(a,n\right)+\sum_{p=0}^{n-1}c_{p}\left(n\right)a^{p}.
\label{fiapro}
\end{equation}

Note that the finite sum comes from the indefinite integrations and it can
be omitted whenever the Laplace transform (\ref{laplace}) fulfills the
following asymptotic condition

\begin{equation}
\lim_{a\rightarrow\infty}S\left(a\right)=0.  \label{eq:asymp}
\end{equation}

In addition, the approximate computation of $S\left(a,n\right)$, and
subsequently of $S\left(a\right)$ from Eq.(\ref{fiapro}), is doable from the
following definition of the Generalized Borel Transform of $%
S\left(a,n\right) $

\begin{equation}
B_{\lambda}\left(s,n\right)\equiv-\int_{0}^{\infty}\exp\left[s/\eta\left(a%
\right)\right]\left[\frac{1}{\lambda\eta\left(a\right)}+1\right]^{-\lambda s}%
\frac{S\left(a,n\right)}{\left[\eta\left(a\right)\right]^{2}}\frac{%
\partial\eta\left(a\right)}{\partial a}da,\quad Re\left(s\right)<0
\label{borel}
\end{equation}
where $\lambda$ is any real and positive number, whereas $\eta$ is defined
so that $1/\eta\equiv\lambda\left(\exp\left(a/\lambda\right)-1\right)$.
Then, the analytical properties of $B_{\lambda}\left(s,n\right)$ allow one
to invert unambiguously Eq.(\ref{borel}) to get

\begin{equation}
S\left(a,n\right)=2\lambda^{2}\left(1-\exp\left(-a/\lambda\right)\right)%
\int_{-\infty}^{\infty}\int_{-\infty}^{\infty}\exp\left[G\left(w,t,a,%
\lambda,n\right)\right]dwdt.  \label{doble}
\end{equation}
being $G\left(w,t,a,\lambda,n\right)$ an involved function depending
explicitly on $H${[}8,9{]}.

Given that the aforementioned expressions (\ref{int}-\ref{doble}) are valid
for any number of  $n\geq 0$, the dominant contribution to the double
integral is computed for $n\gg 1$ by using the steepest descent method$^{13}$
in the variables $t$ and $w$. On the other hand, observe that each value of
the parameter $\lambda $ in Eq.(\ref{borel}) defines a particular Borel
transform. However, the resulting expression for $S\left( a,n\right) $ given
by Eq.(\ref{doble}) does not depend on $\lambda $ explicitly. Consequently,
one can choose the value of this parameter in such a way that it allows one
to solve Eq. (\ref{doble}). Hence, the approximate expression of $S\left(
a,n\right) $ obtained from the saddle point is computed in the limit $%
\lambda \rightarrow \infty $ to finally obtain

\begin{equation}
S_{Aprox}\left(a,n\right)=\sqrt{2\pi}e^{-1/2}\frac{\sqrt{f\left(u_{o}%
\right)+1}}{\sqrt{D\left(u_{o}\right)}}\left[u_{o}\right]^{n+1}H\left(u_{o}%
\right)\exp\left[-f\left(u_{o}\right)\right],  \label{recupapprox}
\end{equation}
where

\begin{equation}
f\left(u_{o}\right)\equiv1+n+u_{o}\frac{d\ln\left[H\left(u_{o}\right)\right]%
}{du_{o}},  \label{fx}
\end{equation}

\begin{equation}
D\left(u_{o}\right)\equiv-u_{o}\,\frac{df\left(u_{o}\right)}{du_{o}}%
\left[1/2+f\left(u_{o}\right)\right]+f\left(u_{o}\right)\left[1+f\left(u_{o}%
\right)\right],  \label{dx}
\end{equation}
and $u_{o}$ is the solution of the following implicit equation

\begin{equation}
u_{o}^{2}a^{2}=f\left(u_{o}\right)\left[f\left(u_{o}\right)+1\right].
\label{gx}
\end{equation}

In summary, the application of this technique consists simply in solving the
implicit equation (\ref{gx}) for $n\gg1$ to obtain the expression of $u_{o}$
and replace it in Eq.(\ref{recupapprox}). Then the $n-$indefinite integrals
appearing into Eq.(\ref{fiapro}) can be solved to get an approximate
solution of $S\left(a\right)$ in the limit $n\rightarrow\infty.$ Observe
that the expressions (\ref{fiapro}-\ref{gx}) appearing in this approach
avoid perturbative expansions and all of them are unambiguously defined.
Moreover, the parameters $n$ and $\lambda$ initially introduced to make the
computation mathematically tractable, are lastly taken away from the
approximate solution for $S\left(a\right).$

Thus, the approximate expression for the Laplace transform appearing in Eq.(%
\ref{newexpgenfun}) is obtained by replacing the function $%
H\left(u\right)=1/\left[\sqrt{u}\sqrt{1+4\sqrt{u}}\right]$ into the
expressions (\ref{fiapro}-\ref{gx}) provided by the GBT. Then, the
expression of $f\left(u_{o}\right)$ given by expression (\ref{fx}) is
replaced into Eq.(\ref{gx}) and $n$ is assumed large to get $%
u_{o}\simeq\left(n+3/4\right)/a$. Then $f\left(u_{o}\right)\simeq n+3/4$ , $%
D\left(u_{o}\right)\simeq\left(n+1/4\right)\left(n+5/4\right)$ and $%
S\left(a,n\right)$ can be approximated as

\begin{equation}
S_{Aprox}\left(a,n\right)\simeq\frac{\sqrt{2\pi}\sqrt{n+5/4}%
\left[n+3/4\right]^{n+1/4}\exp\left[-\left(n+3/4\right)\right]}{2\sqrt{%
\left(n+1/4\right)\left(n+5/4\right)}a^{n+1/4}\sqrt{1+\sqrt{a}/\left[4\sqrt{%
n+3/4}\right]}}  \label{recupapprox2}
\end{equation}

By recognizing the asymptotic expansion of the Gamma function$^{11}$ $\Gamma
\left( n+1/4\right) $ in Eq.(\ref{recupapprox2}), the expression of $%
S_{Aprox}$ can be simplified to get

\begin{equation}
S_{Aprox}\left(a,n\right)\simeq\frac{\Gamma\left(n+1/4\right)}{2a^{n+1/4}%
\sqrt{1+\sqrt{a}/\left[4\sqrt{n+3/4}\right]}}  \label{gapro}
\end{equation}

Then, this expression is replaced into Eq.(\ref{fiapro}), and the coming
expansion

\begin{equation}
\frac 1{\sqrt{1+\sqrt{a}/\left[ 4\sqrt{n+3/4}\right] }}=\sum_{p=0}^\infty 
\frac{L_p^{-1/2}\left( 0\right) }{\left( -4\right) ^p\left( n+3/4\right)
^{p/2}}a^{p/2}\qquad a\rightarrow 0^{+}  \label{exp}
\end{equation}
is utilized, being $L_p^q\left( x\right) $ the Laguerre polynomial{.}$^{11}$
This leads to calculate the non perturbative contribution of $S\left(
a\right) $ as follows

\begin{equation}
S_{NP}\left(a\right)\simeq 
\begin{array}{c}
\lim \\ 
n\rightarrow\infty
\end{array}
\sum_{p=0}^{\infty}\frac{\Gamma\left(n+1/4\right)L_{p}^{-1/2}\left(0\right)}{%
2\left(-4\right)^{p}\left(n+3/4\right)^{p/2}}\left(-\right)^{n}\underbrace{%
\int da\cdots\int da}_{n}\frac{1}{a^{n+1/4-p/2}}  \label{nprecup}
\end{equation}
where

\[
\underbrace{\int da\cdots\int da}_{n}\frac{1}{a^{n+1/4-p/2}}%
=\left(-\right)^{n}\Gamma\left(1/4-p/2\right)/\left[\Gamma\left(n+1/4-p/2%
\right)a^{1/4-p/2}\right]. 
\]

Thus, the expression (\ref{nprecup}) computed in the limit $%
n\rightarrow\infty$ turns into

\[
S_{NP}\left(a\right)=\sum_{p=0}^{\infty}\frac{L_{p}^{-1/2}\left(0\right)%
\Gamma\left(1/4-p/2\right)}{2\left(-4\right)^{p}a^{1/4-p/2}}, 
\]

whose series converges to the following final expression

\begin{equation}
S_{NP}\left(a\right)=\frac{\pi}{4}\left\{ I_{1/4}\left(\frac{a}{32}%
\right)+I_{-1/4}\left(\frac{a}{32}\right)\right\} \exp\left(-\frac{a}{32}%
\right),  \label{nprecupfinal}
\end{equation}

where $I_v\left( z\right) $ is the Bessel function of second class.$^{11}$

Going back to expression (\ref{newexpgenfun}), the solution for the
generating function is completed by solving the second integral appearing
there. It is easily done by using the expansion of the exponential term and
exchanging sum and integral operations. Then, the following solution, valid
for any value of the parameter $a,$ is obtained

\begin{equation}
\int_0^1dx\exp \left[ -ax^2\left( 1-x\right) ^2\right] =\sum_{k=0}^\infty 
\frac{\left( -a\right) ^k\left[ \Gamma \left( 2+2k\right) \right] ^2}{\left(
2k+1\right) ^2\Gamma \left( k+1\right) \Gamma \left( 2+2k\right) }%
=\,_2F_2\left( \left[ \frac 12,1\right] \left[ \frac 34,\frac 54\right]
,-\frac a{16}\right) ,  \label{hypergeom}
\end{equation}
where $_2F_2$ is the well-known Generalized Hypergeometric function.$^{11}$

Finally, the expressions (\ref{nprecupfinal}) and (\ref{hypergeom})
evaluated in terms of the original coupling parameter $g$ provide the
following analytical solution of the generating function in the non
perturbative regime

\begin{equation}
Z_{NP}\left(g\right)=\frac{1}{g}\left[\frac{\pi}{4}\left\{ I_{1/4}\left(%
\frac{1}{64g^{2}}\right)+I_{-1/4}\left(\frac{1}{64g^{2}}\right)\right\}
\exp\left(-\frac{1}{64g^{2}}\right)\right.  \label{finalexpression}
\end{equation}

\[
\left.+{}_{2}F_{2}\left(\left[\frac{1}{2},1\right]\left[\frac{3}{4},\frac{5}{%
4}\right],-\frac{1}{32g^{2}}\right)\right] 
\]

The test of the accuracy provided by the GBT result is presented in Fig. 1.
The relative error with respect to the exact solution (\ref{fungen}) (solid
line) is less than 5\% for values of $g$ larger than $100.$ It shows that
the GBT is a very efficient tool to capture non perturbative contributions
of this model.

In summary, expressions (\ref{regborel22}), (\ref{regborel1}) and (\ref
{finalexpression}) for the generating functional of the zero-dimensional $%
\Phi^{4}$ theory with degenerate minima in the non perturbative regime of
the coupling parameter $g$ were obtained by application of the Borel
resummation and GBT techniques. Models like this are mainly utilized for
studying the physics of barrer penetration and tunelling effects in systems
which classical minima are connected so that the symmetry between them is
not spontaneously broken.

It was found that the genuine origin of the main non perturbative
contributions of this model is not the Borel ambiguity. This non analyticity
of the Borel sum is due to the non alternating character of the coefficients
in the expansions in powers of $g.$ In fact, this behavior is a direct
consequence of the degeneracy of the ground state and the presence of
inevitable contributions, which cannot be captured from Perturbation Theory.
It implies that for reconstructing the generating function one needs
additional information which is absent in Perturbation Field Theory. Indeed,
this approach has to introduce a parameter to be afterward fitted in order
to quantify non perturbative contributions.

On the other hand, the GBT was established to be a very efficient tool to
capture non perturbative contributions of this model. The overall goal of
this technique comes from the fact that it involves expressions
unambiguously defined which preserve the complete information about the
coupling parameter. Indeed, the computation of the analytical solution is
doable due to the help of auxiliary parameters which do not appear in the
final results. The main role of the real parameter of the model is simply to
control the localization of the saddle point $u_{o}$ and consequently the
behavior of the resulting final solution.

The author thanks M. Pettitt and the Institute for Molecular Design, Houston
for the generous and warm hospitality extended to him. He also thankful H.
Fanchiotti, Luis Epele and C.A. Garcia Canal for their advice and helpful
discussions.

\section*{References}

{\small 1. J.C. Le Guillou and J. Zinn-Justin,} \emph{\small Large Order
Behaviour of perturbation Theory}{\small , (North-Holland, amsterdam 1989);
Jean Zinn-Justin,} \emph{\small Quantum Field theory and critical phenomena}%
{\small , 3rd edition (Oxford University Press, New York 1995); C. Itzykson,}
\emph{\small Quantum Field Theory}{\small , (McGraw-Hill 1980); Hagen
Kleinert,} \emph{\small Path integrals in Quantum Mechanics Statistics and
Polymer Physics}{\small , 2nd edition (World Scientific, Singapore 1995);
Yu.A. Simonov,} \emph{\small Lectures at 13th Indian Summer School}{\small ,
(Prague, Czech Replublic 2000) (hep-ph/0011114).}

{\small 2. C.M. Bender and S.A. Orszag,} \emph{\small Advanced Mathematical
Methods for scientistic and Engeneers}{\small , (McGraw-Hill 1978);}

{\small 3. A.A. Penin, A.A. Pivovarov, Phys. Lett} \textbf{\small B401}%
{\small , 294 (1997).}

{\small 4. L.S. Brown, L.G. Yaffe and C. Zhai, Phys. Rev.} \textbf{\small D46%
}{\small , 4712 (1992) ; G.V. Dunne and T.M. Hall, Phys. Rev.} \textbf%
{\small D60}{\small , 065002 (1999) ; M. Neubert, Phys. Rev.} \textbf{\small %
D51}{\small , 5924 (1995) ; G. Grunberg, Phys. lett.} \textbf{\small B} 
{\small 95, 70 (1980).}

{\small 5. G.N. Hardy,} \emph{\small Divergent Series}{\small , (Oxford
Univ. Press 1949).}

{\small 6. M. Beneke, Phys.Rept.}\textbf{\small 317}{\small , 1 (1999).}

{\small 7. L.N.Epele, H. Fanchiotti, C.A. Garcia Canal and M. Marucho, Nucl.
Phys.} \textbf{\small B583}{\small , 454 (2000); L.N.Epele, H. Fanchiotti,
C.A. Garcia Canal and M. Marucho, Phys. Lett.} \textbf{\small B523}{\small ,
102 (2001).}

{\small 8. G.A. Carri and M. Marucho, J. Math. Phys.} \textbf{\small 44}%
{\small , 6020 (2003).}

{\small 9. L.N.Epele, H. Fanchiotti, C.A. Garcia Canal and M. Marucho, Phys.
Lett.} \textbf{\small B556}{\small , 87 (2003).}

{\small 10. E.R. Caianiello, G. Scarpetta, N. Cim.} \textbf{\small 22A}%
{\small , 448 (1974); ibid Lett. N.Cim} \textbf{\small 11}{\small , 283
(1974); H.G. Dosh, Nucl. Phys.} \textbf{\small 96}{\small , 525 (1975); R.J.
Rivers, J. Phys.} \textbf{\small A13}{\small , 1623 (1980);
A.P.C.Malbouisson, R. Portugal and N.F. Svaiter, Phys.} \textbf{\small A292}%
{\small , 485 (2001); C.M. Bender, S.Boettcher and L. Lipatov, Phys. Rev.} 
\textbf{\small D46}{\small , 5557 (1992); J. Zinn-Justin, J. Math. Phys.} 
\textbf{\small 22}{\small , 511 (1981); C. Bachas, C. de Calan, and P.M.S.
Petropoulos, J. Phys. A: Math Gen.} \textbf{\small 27}{\small , 6121 (1994);
C.F. Baillie, W. Janke, D.A. Johnston and P. Plechac, Nucl. Phys.} \textbf%
{\small B450}{\small , 730 (1995); B. Derrida, Phys. Rev.} \textbf{\small B24%
}{\small , 2613 (1981); M. Aizeman, J.L. Lebowitz, D. Ruelle, Comm. Math.
Phys.} \textbf{\small 112}{\small , 3 (1987).}

{\small 11. I.S. Gradshteyn and I.M. Ryzhik,} \emph{\small Table of
Integrals, Series, and Products} {\small (Academic Press, New York 2000); M.
Abramowitz and I. Stegun,} \emph{\small Handbook of Mathematical Functions} 
{\small (Dover, New York 1970); H. Bateman,} \emph{\small Higher
Transcendental Functions} {\small (Mc Graw Hill, New York 1953); A.
Nikiforov y V.B. Uvarov,} \emph{\small Special functions of mathematical
physics}{\small ,third edition (Cambridge 1966); A. Prudnikov, O. Marichov
and Yu. Brychlov,} \emph{\small Integrals and series}{\small , (Newark, NJ,
Gordon and Breach 1990).}

{\small 12. M. Pindor, hep-th/9903151; I. Caprini and M. Neubert, JHEP} 
\textbf{\small 9903}{\small , 007 (1999).}

{\small 13. E.T. Copson,} \emph{\small Asymptotic expansions}{\small ,
(Cambridge 1965); Norman Bleistein y Richard Handelsman,} \emph{\small %
Asymtotic expansions of integrals}{\small , (Dover, New York 1986); H.
Jeffrey and B.S. Jeffrey,} \emph{\small Methods of Mathematical Physics} 
{\small (Cambridge University Press 1966).}

\newpage

\begin{figure}[tbp]
\includegraphics[ scale=0.65, angle=270]{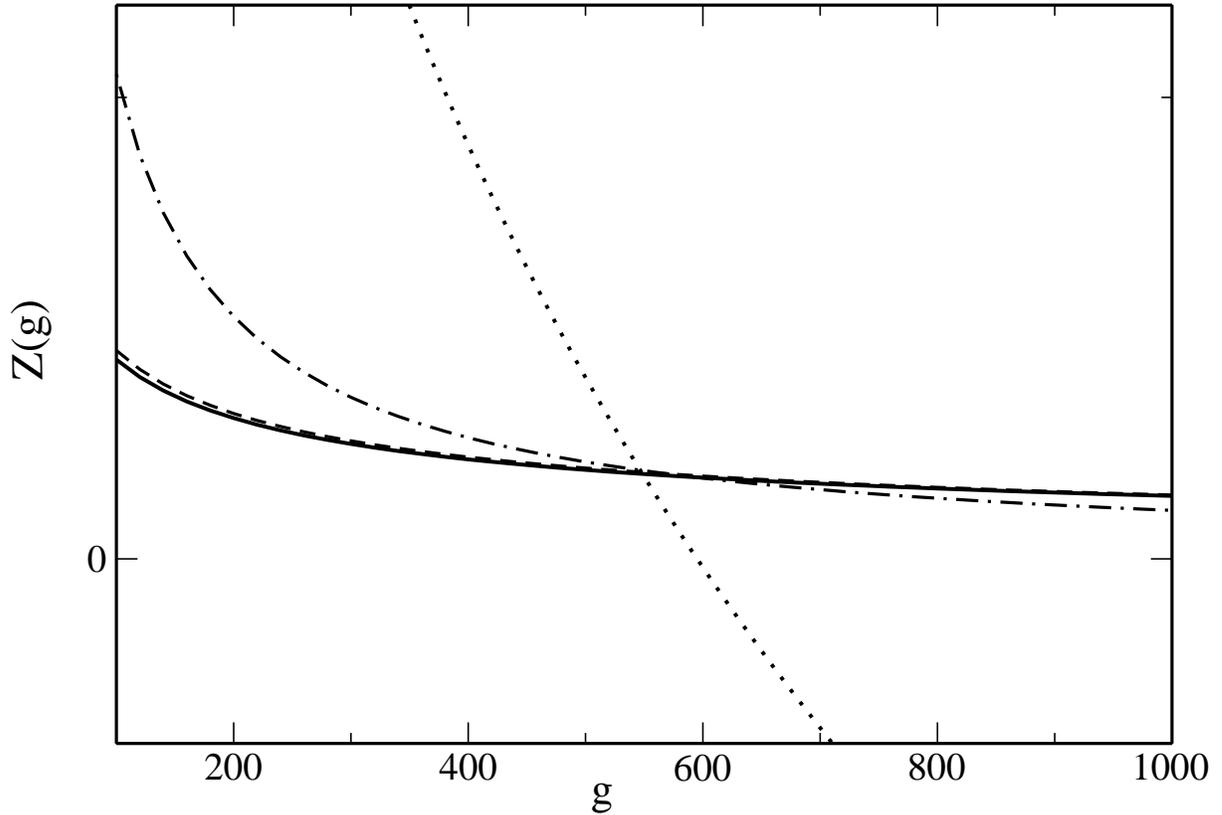}
\caption{Generating function $Z\left(g\right)$ as a function of the coupling
parameter $g$. The solid line corresponds to the numerical evaluation of the
exact solution (\ref{fungen}) and the dashed line to the GBT solution (\ref
{finalexpression}). Dotted-dashed and dotted lines correspond to Borel
resummation solutions (\ref{regborel1}) and (\ref{regborel22}) respectively.
The value of $C$ minimizing the deviation of these approaches with respect
to the exact solution (\ref{fungen}) (solid line) was found to be equal to $%
-47.313$ and $3.572$ respectively. They were computed by the least square
method in the range of values of $g$ given by the interval $[100,1000].$}
\end{figure}

\end{document}